\documentclass[prd,twocolumn,superscriptaddress,nofootinbib,showpacs,showkeys,floatfix, groupedaddress]{revtex4-1}

\pdfoutput=1

\setcitestyle{square,numbers}

\usepackage{graphicx} % for figures
\usepackage{amssymb,amsmath}
\usepackage{adjustbox}
\usepackage{enumitem}
\usepackage{hyperref}

\usepackage{soul}
\setstcolor{red}

\usepackage[usenames]{color}
\usepackage[dvipsnames]{xcolor}

\usepackage{pgfplots}
\pgfplotsset{
  compat=1.3, %this makes the y axis label move closer to the plot
}

\ifnum\pdfstrcmp{\jobname}{paper-lgimg}=0
    
\else
    
\fi

\pgfmathsetmacro\TrimWidth{0.23}
\pgfmathsetmacro\TrimWidthEMTop{0.28}
\pgfmathsetmacro\TrimWidthForGeneric{0.29}

\begin{document}

\title{What does a binary black hole merger look like?}

\newcommand{\Caltech}{\affiliation{Theoretical Astrophysics 350-17,
    California Institute of Technology, Pasadena, California 91125,
    USA}} %
\newcommand{\Cornell}{\affiliation{Center for Radiophysics and Space
    Research, Cornell University, Ithaca, New York 14853, USA}}%
\newcommand{\MIT}{\affiliation{MIT Kavli Institute,
    Massachusetts Institute of Technology, Cambridge,
    Massachusetts 02139, USA}}%

\author{Andy Bohn}
\email[Lensing group email: ]{lensing@black-holes.org}
\Cornell
\author{William Throwe} \Cornell
\author{Fran\c{c}ois H\'{e}bert} \Cornell
\author{Katherine Henriksson$^\dagger$} \Cornell
\author{Darius Bunandar} \Caltech \MIT
\author{Mark A.~Scheel} \Caltech
\author{Nicholas W.~Taylor} \Caltech

\date{\today}

\begin{abstract}
  We present a method of calculating the strong-field gravitational lensing
  caused by many analytic and numerical spacetimes.
  We use this procedure to
  calculate the distortion caused by isolated black holes and by numerically
  evolved black hole binaries.  We produce both
  demonstrative images illustrating
  details of the spatial distortion and realistic images
  of collections of stars taking both lensing amplification and redshift into
  account.  On large scales the lensing from inspiraling binaries resembles
  that of single black holes, but on small scales the resulting images show
  complex and in some cases self-similar structure across different angular
  scales.
\end{abstract}

\maketitle
\renewcommand{\thefootnote}{\fnsymbol{footnote}}
\footnotetext[2]{Current affiliation: Google Inc., 747 6th St S,
Kirkland, WA 98033, USA}
\renewcommand{\thefootnote}{\arabic{footnote}}

\section{Introduction}
\label{sec:introduction}

Black holes are the most compact gravitating objects in the
  universe, with
  such strong gravitational fields that
  not even light can escape them. In the vicinity of a
  black hole,
  light rays can be very strongly deflected from a straight-line
  path, sometimes orbiting around the black hole before continuing on their
  way.
  It is now well-known that the bending of light by massive objects
  like galaxy clusters can create
  brightness amplification~\cite{Refsdal1964},
  deformed images, or even multiple images~\cite{Blandford1992}
  of background objects such as quasars.
  These signatures have
  so far only been directly observed in
  cases where the deflection of light is very slight,
  up to approximately $11$ arc seconds~\hbox{\cite{Inada2003, Inada2006}}.
  However, here we are interested
  in the lensing effects associated with
  much more extreme bending of light near
  single or binary black holes,
  where the deflection angle is unbounded.

The lensing effects near general-relativistic bodies were
first studied in the 1970s,
with Cunningham and Bardeen~\cite{Cunningham1972}
looking at a star on an orbit in a Kerr spacetime, and
Luminet~\cite{Luminet1979}
studying an accretion disk around a Schwarzschild black hole.
More recently, open-source codes such as
GYOTO~\cite{Vincent2011} and GeoViS~\cite{Muller2014}
have produced images of lensing in the
neighborhood of various compact objects.
While the lensing caused by an isolated black hole has been understood
analytically, the case of lensing by a binary black hole (BBH) is much more
challenging because of
the difficulty of solving for the geometry of the spacetime.
With some arguably unrealistic assumptions
(e.g.,~two maximally charged black holes in static equilibrium),
analytic solutions can be found and
subsequently used for lensing~%
\cite{Majumdar1947,Papapetrou1947,Kastor1993,Nitta2011,Yumoto2012,Muller2014}.

\begin{figure}
\centering
  \includegraphics[width=.9\columnwidth]{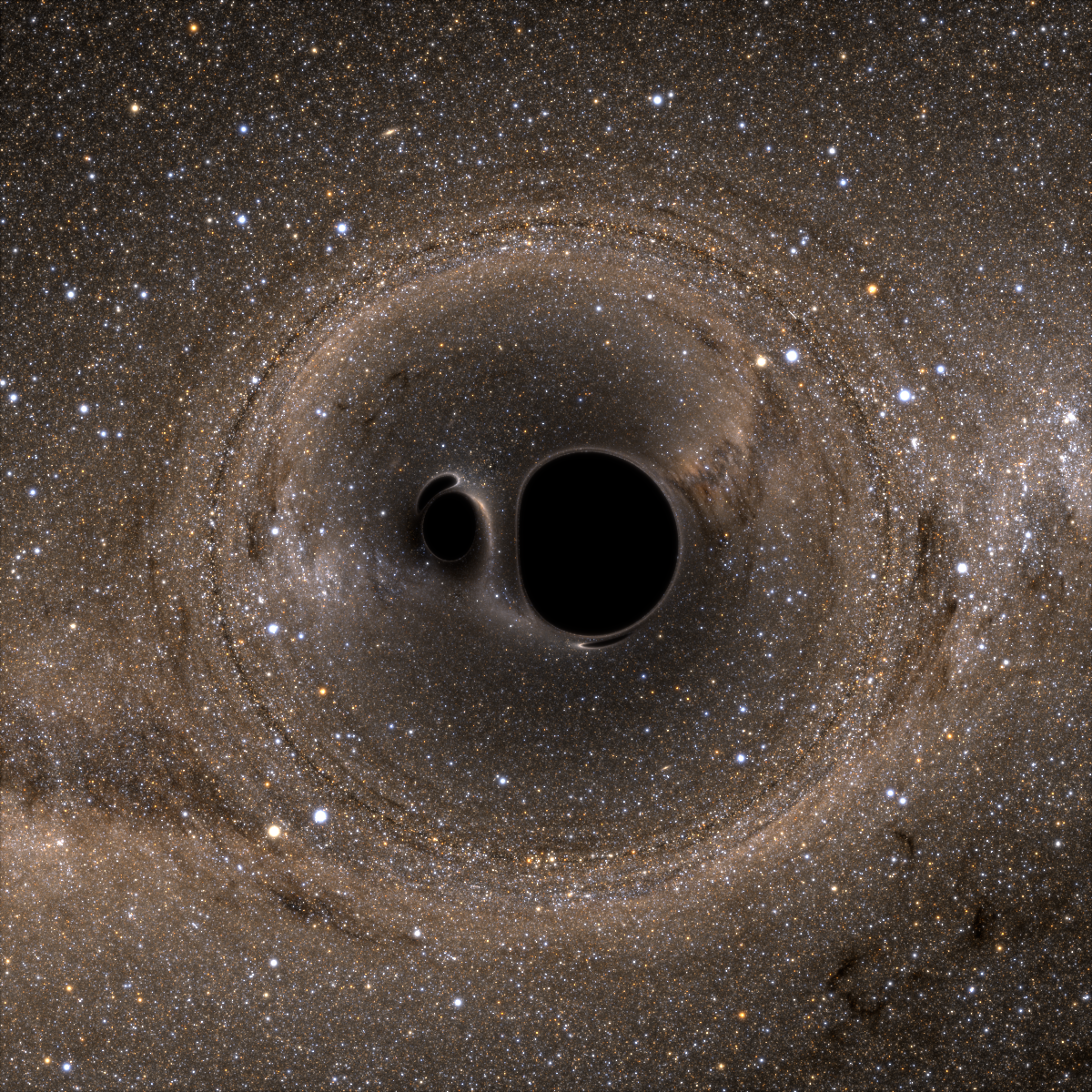}
  \caption{
    A pair of black holes that are about to merge, with the Milky Way
    visible in the background.
    Supplementary images and movies can be found at \cite{LensingWebsite}.
  }
  \label{fig:Stars}
\end{figure}

For astrophysically relevant
  binaries, however, we must instead
  rely on numerical solutions.
  Solving these binary spacetimes numerically to high accuracy
  has been possible for the last decade
  (see~\cite{Centrella:2010mx, Pfeiffer:2012pc} for a review),
  motivated by the need to provide
  gravitational-wave templates used by experiments
  such as LIGO, VIRGO, and KAGRA to make detections.
  By using
  the spacetimes computed in such simulations,
  we gain the ability to solve for the lensing effects in BBH systems.

In this paper, we focus on the question of what an observer
  in the vicinity of a BBH would actually \emph{see} as
  the black holes orbit, spiral inward, and merge, with an example
  shown in figure~\ref{fig:Stars}.
  This is in contrast to most BBH
  visualizations, in which the
  positions or horizons
  of the two black holes are simply shown as a function of time
  in some coordinate system.
  We instead
  compute the paths of light rays that
  enter the observer's eye or camera to find what
  would actually
  be seen.
  Furthermore, this path must be computed in the fully time-dependent spacetime,
  as the orbital velocities for a black-hole binary are typically large enough
  that the system cannot be approximated as
  time-independent during the time taken
  by the photons to travel across it.

Because the black holes themselves do not emit light (we ignore
  Hawking radiation, which is significant only for microscopic black
  holes), the observer would see nothing unless there is some
  additional light source. For illustrative purposes, we will take
  an artificial background ``painted on'' at
  infinity (figure~\ref{fig:CutSphere})
  as the light source for most of our examples;
  this will allow
  us to study in detail where each light ray originates.

We begin by describing the problem setup and the methods that we use to
generate lensing images in section~\ref{sec:methods}.
In section~\ref{sec:results} we show images of lensing by
single and binary black holes,
and we then conclude in section~\ref{sec:conclusion}.

\section{Methods}
\label{sec:methods}

We set up the problem with our black hole(s)
near the center of our chosen coordinate system.
While any physical
system representable by a spacetime metric can be used,
we specialize in this paper to single and binary black holes.
The observer (henceforth taken to be a
camera) can be located anywhere in the space
and is typically chosen to look towards the origin.
A sphere with our light source
encloses the black hole(s) and camera, infinitely far away.

To recreate the image taken by the camera in this configuration, we
must find the properties of the light that arrives at each point on the
camera's image plane.
A na{\"i}ve approach would be to trace all possible light rays
(i.e.,~null geodesics) emanating from
the light source to determine which rays reach the camera
and from what directions they arrive,
but this is computationally infeasible.
A more efficient approach is to
reverse the problem by tracing light rays away
from the camera and \emph{backwards} in time
(the computer graphics community calls this a ray-casting algorithm).
This method identifies the origin of any
light ray that illuminates the camera, from which we infer
the color and intensity of the corresponding
photons as detected by the camera.
When black holes are present, some of the null geodesics traced
backwards in time from the camera may approach arbitrarily
close to an event horizon as
$t\to -\infty$; these geodesics correspond to dark image regions.

In what follows we describe how the light rays are traced
from the camera using the
geodesic language from general relativity. We show how we
initialize these
geodesics based on camera parameters such as position and viewing angle.
Finally,
we show how the origin of each light ray is determined and describe how the
simulated image is constructed.

\subsection{Geodesic tracing}
\label{sec:geodesic-evolution}

Our code can trace geodesics independently
through either numerical or analytic metric data.
It is common for numerical simulations
to use the $3+1$ decomposition~\cite{ADM},
so we
express the metric in the form
\begin{equation}
  \label{eqn:Decomposition}
  ds^2 = -\alpha^2 dt^2 + \gamma_{ij}(dx^i + \beta^i dt) (dx^j + \beta^j dt),
\end{equation}
where $\alpha$ is the lapse function, $\beta^i$ is the shift vector,
and $\gamma_{ij}$ is the spatial metric.\footnote{
  Our convention is that
  Greek indices, as in $x^\lambda$,
  denote temporal or spatial components, while Latin
  indices, as in $x^i$, denote only spatial components.}
We obtain numerical data from simulations performed using the
Spectral Einstein Code
(SpEC)~\cite{SpECwebsite,
SXSWebsite, Szilagyi:2009qz, Hemberger:2012jz, SXSCatalog}.
The geodesics are traced by evolving a solution to
the geodesic equation
\begin{equation}
\label{eq:nullGeodesic}
  \frac{d^2 x^\lambda}{d\tau^2} + \Gamma^{\lambda}_{\phantom{\lambda}\mu \nu} \frac{dx^{\mu}}{d\tau} \frac{dx^{\nu}}{d\tau} = 0,
\end{equation}
where $x^\lambda$
is the four-position of the geodesic, $\tau$ is an affine
parameter, and $\Gamma^{\lambda}_{\phantom{\lambda}\mu \nu}$ are the
Christoffel symbols describing the effective force
caused by spacetime curvature.

To facilitate the numerical geodesic evolution,
we split this second-order differential equation into two
first-order differential equations using an intermediate,
momentum-like variable such as $p^\lambda = dx^\lambda / d\tau$.
As we have some freedom in the definition of this momentum variable,
we look for one
that helps to
minimize computational time and numerical errors
when evolving through spacetimes with black holes.

We initially explored using the variable
$p_\lambda = g_{\lambda \kappa} p^\kappa$ from Hughes \textit{et al.}~\cite{Hughes1994},
along with converting
the evolution equations from affine parameter $\tau$ to the coordinate time
$t$ of SpEC evolutions through the use of $p^0 = dt/d\tau$.
Although the resulting evolution equations are
concise and have no time
derivatives of metric variables,
the variables $p^0$ and $p_i$ grow
exponentially near black hole horizons in
typical coordinate systems used by SpEC
simulations.  This forces our time-stepper to take
prohibitively small steps in order to achieve the desired accuracy.

We therefore choose a momentum variable slightly different than $p_\lambda$
to mitigate this time-stepping problem.
Null geodesics satisfy \hbox{$p \cdot p = 0$},
which can be rewritten as
\hbox{$p^0=\alpha^{-1}(\gamma^{ij} p_i p_j)^{1/2}$} using the
metric~\eqref{eqn:Decomposition}.
This expression shows that $p^0$ and $p_i$ scale similarly, so we can eliminate
the exponential behavior of these variables by evolving the ratio.
Our intermediate variable thus becomes
\begin{equation}
  \label{eqn:PiDefinition}
  \Pi_i \equiv \frac{p_i}{\alpha p^0} = \frac{p_i}{\sqrt{\gamma^{jk} p_j p_k}},
\end{equation}
where we also divide by $\alpha$ to reduce the number of terms
in the resulting evolution
equations.
Using $\Pi_i$ and the $3+1$ decomposition~\eqref{eqn:Decomposition},
we can express the geodesic equation~\eqref{eq:nullGeodesic}
in the form
\begin{equation}
\label{eq:qpEvolution}
\begin{split}
  \frac{d \Pi_i}{dt} = {}& -\alpha_{,i}
    + (\alpha_{,j} \Pi^j - \alpha K_{jk} \Pi^j \Pi^k) \Pi_i \\
    & + \beta^k_{\phantom{k} ,i} \Pi_k
    - \frac{1}{2}\alpha \gamma^{jk}_{\phantom{jk},i} \Pi_j \Pi_k, \\
  \frac{ dx^i}{dt} = {}& \alpha \Pi^i - \beta^i,
\end{split}
\end{equation}
where $K_{jk}$ is the extrinsic curvature (see, e.g.,~\cite{ADM}) and
$\Pi^i$ is defined via the inverse spatial metric as
$\Pi^i \equiv \gamma^{ij} \Pi_j$.
Note that the geodesic equation
consists of four second-order equations, yet we only
have three pairs of coupled first-order equations in~\eqref{eq:qpEvolution}.
Because we are evolving a normalized momentum~\eqref{eqn:PiDefinition},
we have lost information about $p^0$ during evolution.
Compared to Hughes \textit{et al.}~\cite{Hughes1994},
we have introduced a time derivative of the
three-metric inside $K_{jk}$, but we
have significantly sped up the evolution near
black holes by removing the exponential growth of $p^0$ and $p_i$.

The equations in~\eqref{eq:qpEvolution} are similar to those in~(28) of
Vincent \textit{et al.}~\cite{Vincent2012}.
In fact our intermediate evolution variable $\Pi_i$ is related to their
variable $V^i$ by the three-metric,
such that $\Pi^i = V^i$.
But our~\eqref{eq:qpEvolution} has a
reduced number of both temporal and spatial derivatives of metric
quantities compared to Vincent's~(28).

During the backwards-in-time geodesic evolution,
many geodesics are traced until they
are far from the strong-field region,
but some are traced until they
encounter a black hole.
These latter geodesics
slowly converge towards the black hole's event horizon,
but as they can in principle
be evolved indefinitely, we
need some way of identifying them in finite time.
We do this by monitoring $p^0$ for each
geodesic,
which (as discussed above)
grows large near black hole horizons.
Since our evolution equations~\eqref{eq:qpEvolution} do not evolve $p^0$,
we
must evolve another equation to keep track of it.  However, we would
still like to avoid the exponential growth of $p^0$
near the horizon.
This can be accomplished by evolving the logarithm of $p^0$.
As was done in~\eqref{eqn:PiDefinition}, we multiply
$p^0$ by the lapse to reduce the number of terms in the resulting
equation, which gives the evolution
variable $\ln(\alpha p^0)$. This leads
to the evolution equation
\begin{equation}
\label{eq:p0Evolution}
\begin{split}
  \frac{d \ln(\alpha p^0)}{dt} = {}& -\alpha_{,i} \Pi^i
    + \alpha K_{ij} \Pi^i \Pi^j.
\end{split}
\end{equation}
When $p^0$ becomes too large, signaling a large energy, we flag the
geodesic
as originating from the black hole and we stop evolving it.

The remaining geodesics are those that originate
from infinity,
so we need to determine the ($\theta$,$\phi$) location at infinity where
they come from. In section~\ref{sec:image-generation}, we will need
the gravitational
redshift $z$ of each photon,
which can be calculated from the ratio of the photon's
energy at the two ends of its
trajectory via
\begin{equation}
  1+z = \frac{E_\infty}{E_\text{camera}},
  \label{eq:Redshift}
\end{equation}
where $E_\infty$ is the photon's energy at infinity, and $E_\text{camera}$ is the
photon's energy as measured by the camera.
Therefore we will need to compute the energy that each photon would have
at infinity.
In practice, these geodesics are traced
backwards in
time until they reach a large distance $R$ from the black hole(s),
chosen so that the metric at $R$ is equal to the flat space
metric within about a percent error. We use the approximation
that the metric is exactly flat for $r>R$.
Under this approximation,
the geodesic's
direction and $p^0$ at infinity are the same as
at $R$.  The direction is used to calculate a
($\theta$,$\phi$) location on the sky, while $p^0$ is the
photon's energy at infinity, $E_{\infty}$.

\subsection{Initial data}
\label{sec:initialdata}
Here we outline how we initialize our geodesic
evolution variables.
Because the geodesics are traced
away from the camera, backwards
in time,
we
initialize
each geodesic's evolution variables to their values
at the camera.
We have seven variables to set: three each for the initial position
and momentum in~\eqref{eq:qpEvolution},
and one for the initial redshift in~\eqref{eq:p0Evolution}.

\begin{figure}
\centering
%  \adjincludegraphics[trim={0.02\width} {0.06\height} {0.02\width} {0.06\height},%
%    clip, width=0.5\columnwidth]{pinhole.pdf}
  \adjincludegraphics[trim={0.02\width} {0.06\height} {0.02\width} {0.06\height},%
    clip, width=0.5\columnwidth]{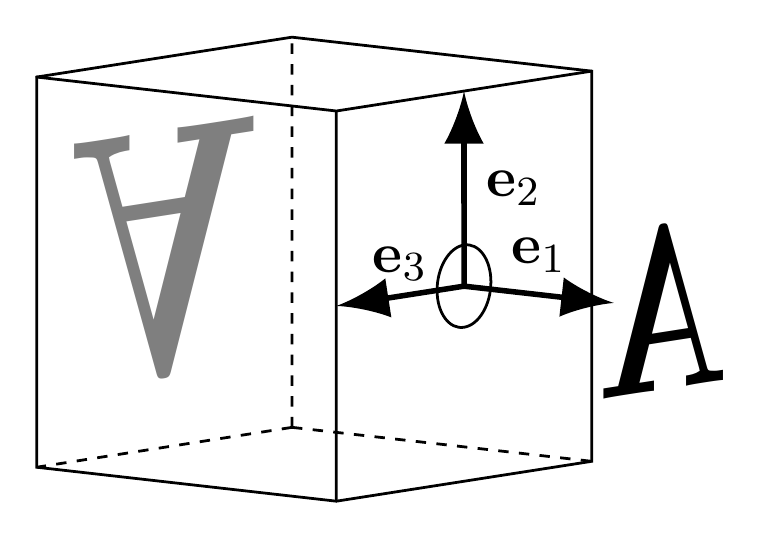}
  \caption{Illustration of a pinhole camera
    in its rest frame
    with the three vectors
    $e_1$, $e_2$, and $e_3$ that describe its orientation.
    The inverted
    letter ``A'' demonstrates the optical properties
    of the camera,
    which we correct for in the images we generate.
  }
  \label{fig:tetrad}
\end{figure}

The initial position for every geodesic
is simply the camera's position.
The initial momentum, however, is different for each
geodesic and is dependent on
the angle at which it enters the camera.
We express the momenta in terms of an orthonormal tetrad defined as
\begin{description}[labelwidth=!,align=right]
\item[$e_0$] The camera's four-velocity, a timelike vector.
For stationary
cameras $e_0 \propto (1,0,0,0)$;
\item[$e_1$] The direction in which the camera is pointing;
\item[$e_2$] The ``upward'' direction for the camera;
\item[$e_3$] The ``rightward'' direction for the camera.
\end{description}
The four-vectors $e_1$, $e_2$, and $e_3$ are all
spacelike, and
their orientations in the camera's reference frame
are illustrated in figure~\ref{fig:tetrad}.

In order to specify this tetrad,
we give guesses for
the vectors $e_0$, $e_1$, and $e_2$, with
the condition that the guessed
time components of $e_1$ and $e_2$ must be zero.
We then apply the Gram-Schmidt process to the sequence $e_0$,
$e_1$, and $e_2$ to transform
these vectors into an orthonormal
set.
The final vector, $e_3$, is found by calculating the
generalized cross product of the other three; explicitly,
\begin{equation}
 {e_3}_\rho=\epsilon_{\lambda\mu\nu\rho}{e_0}^\lambda{e_1}^\mu{e_2}^\nu,
\end{equation}
where $\epsilon_{\lambda \mu \nu \rho}$ is the Levi-Civita tensor
(see~\cite[p.~202]{MTW} for more details).

Given the four orthonormal unit vectors, we can construct a null vector
$\xi$ tangent to the geodesic
that enters the camera from a given direction.
The vector $\xi$ will be proportional
to the four-momentum of a photon following the geodesic; that is,
$p = q \xi$ for some
positive
constant
$q$.  We define $\xi$ by
\begin{equation}
\begin{split}
\xi^{\lambda}_{\phantom{\lambda} (a,b)} =
  C e_0^{\phantom{0} \lambda} - e_1^{\phantom{1} \lambda}
    &- [(2 b - 1) \tan(\alpha_v /2)] e_2^{\phantom{2} \lambda} \\
    &- [(2 a - 1) \tan(\alpha_h / 2)] e_3^{\phantom{3} \lambda},
\label{eq:photonmomentum}
\end{split}
\end{equation}
where $a,b \in [0,1]$ give the ray's arrival direction in terms of
fractions of the image's horizontal and vertical lengths, respectively, and
$\alpha_{v},\alpha_{h}$ are
the angular sizes of the camera aperture
(field of view angles) in the vertical and horizontal directions.
For the sign convention
chosen in~\eqref{eq:photonmomentum}, $(a,b) = (0,0)$
corresponds to a photon seen at the bottom left
  corner of the image.
We find $C$ by requiring that $\xi$
is null, i.e.,
$\xi \cdot \xi = 0$:
\begin{equation}
C = \sqrt{1 +
  (2 b - 1)^2 \tan^2(\alpha_v /2) +
  (2 a - 1)^2 \tan^2(\alpha_h /2)}.
\end{equation}
We then use the metric to lower the index on $\xi$, and we compute
the initial value of our evolution variable $\Pi_i$ using
$\Pi_i=p_i/(\alpha p^0)=\xi_i/(\alpha \xi^0)$.
Note that $\Pi_i$ is independent of
the proportionality constant $q$ relating $\xi$ and the actual
photon momentum $p$; physically,
this is because the
photon trajectory is independent of the photon energy. The only place where
$q$ enters is in the initial value of
$\alpha p^0$ in~\eqref{eq:p0Evolution}.
We fix the value of $q$ by demanding that the
energy of the photon in the frame of the camera be unity when the photon
strikes the camera, so $E_\text{camera} = 1$ in~\eqref{eq:Redshift}.

\subsection{Image generation}
\label{sec:image-generation}

  We create our image of the physical system by dividing
  the image plane into
  rectangular regions corresponding to the pixels of the output image
  and assigning an appropriate color to each region.
  Because each region has an extended size, there is no single source
  point we can look at to obtain its color, so we must adopt some
  prescription for assigning a single color to each
    pixel.
  We use two different prescriptions, based on the nature of the
  light source illuminating the system.

\begin{figure}
\centering
  \includegraphics[width=.5\columnwidth]{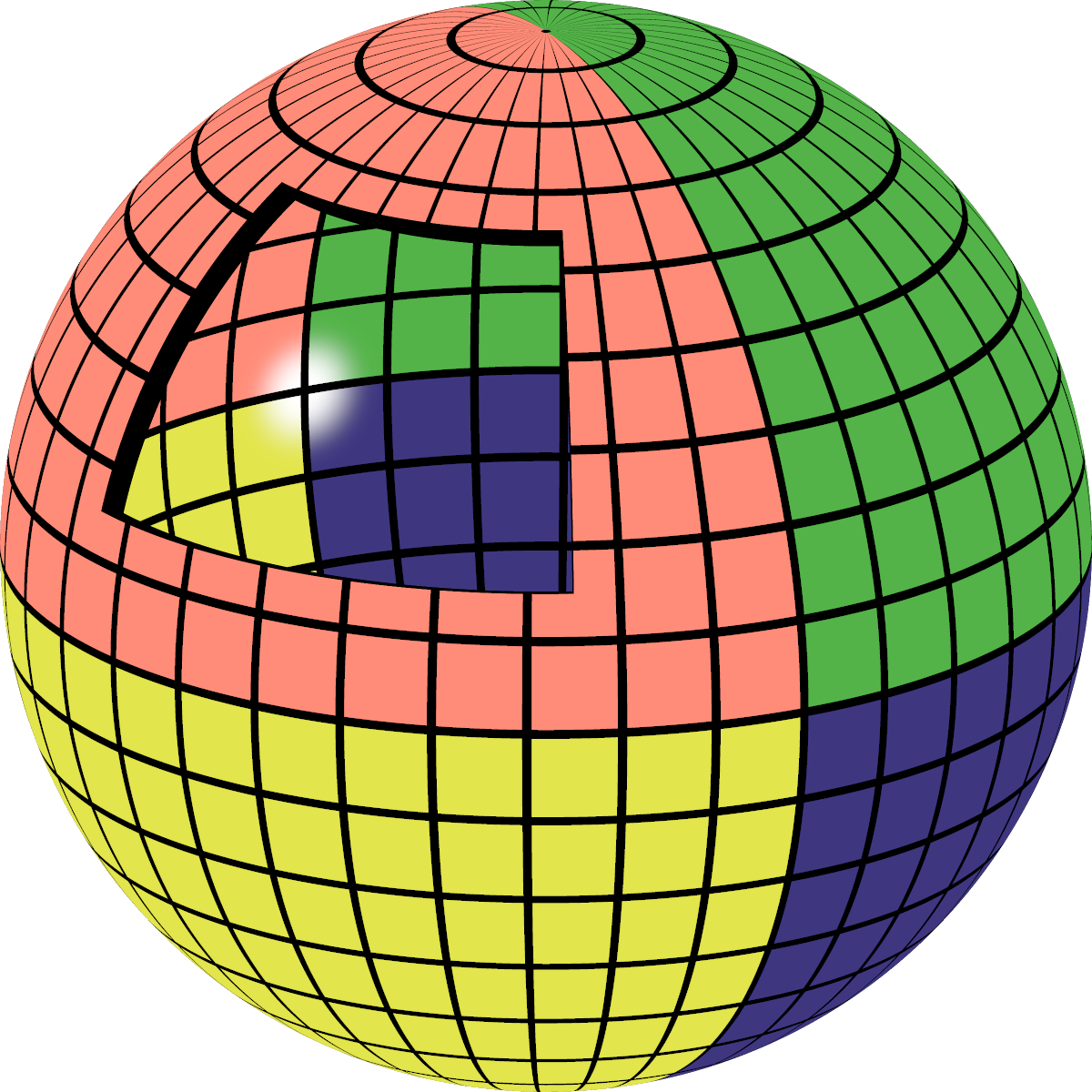}
  \caption{
      An illustration of our artificial background grid ``painted on''
      a sphere at infinity.
      This background is used for all the images with a grid in this paper.
      In the figure, we
      cut a window out of the sphere to show the inside.
      In addition to four colors differentiating the regions of the sphere,
      we include a white reference spot in the direction in which the camera
      is pointing.
    }
  \label{fig:CutSphere}
\end{figure}

  For extended sources, such as the artificial grid in
  figure~\ref{fig:CutSphere},
  we use a subpixel sampling method. On each pixel we construct
  an evenly spaced
  grid of points, and at each of these points we determine where
  incident light rays originate,
  either from one of the holes or a location at infinity.
  We assign a color to each grid point based on that of the
  corresponding source point;  the color of the pixel is then the
  average of these.
  We find that a grid of $4\times4$ sample points gives
  sufficiently smooth
  images without too much computational cost.
  For these images, we neglect the effects of redshift and focus
  on the spatial distortions.

  To create more astronomically relevant images,
  we wish to
  use a collection of point sources (i.e.,~stars) as our
  illumination.  In this case we
  cannot determine a pixel's
  color using sampling, but must instead sum the
  contributions from all the point sources contributing light there.  For our
  list of sources, we use about $3.4\times10^8$ stars from the Two Micron All
  Sky Survey (2MASS)~\cite{Skrutskie:2006}.  To simplify computations, we
  approximate each star as a thermal source with temperature and brightness
  determined by fitting to the photometric information in the catalog.  When we
  calculate the contribution of each star to the light arriving at the camera,
  we must account not only for its properties as a light source, but also for
  the effects of
  the spacetime curvature encountered by the photon.
  These effects come in two forms.  First, the observed energies of
  photons at the camera will be modified by redshift effects, changing sources'
  apparent brightnesses and temperatures.  Second, the spatial convergence or
  divergence of nearby geodesics produces an overall adjustment to each
  source's apparent brightness without affecting its spectrum.
  Both of these effects are discussed in detail in
  Mollerach and Roulet~\cite{Mollerach2002}.
  After we have drawn the entire image in this manner, we convolve it with a
  blurring function to make the stars more visible.  This has the effect of
  transforming each star into a fuzzy circle with size dependent on its
  brightness.

  The result of this scheme can be seen in figure~\ref{fig:Stars},
  which shows the BBH image from figure~\ref{fig:GenericTop}
  in front of a background of stars.
  Note that by generating our
  starfield images from a catalog of point sources,
  we obtain a substantially more realistic image than would be
  generated by applying the lensing deformation to a raster
  image of the unlensed Milky Way stars.
  In such a raster image, each star is usually represented
  (whether as a result of camera optics or software rendering)
  as a blurred circle whose area depends on the star's brightness.
  These circles are typically hundreds of arc seconds wide,
  and therefore lensing distortions
  applied to the image tend to produce stars that appear as smeared ellipses.
  In contrast, the angular sizes of real stars are many orders of magnitude
  smaller, so we expect them to remain as
  unresolved points under all but the
  most extreme lensing magnifications.
  These unresolved points can then be rendered
  as previously described,
  giving stars that better portray what an observer would actually see
  (as in figure~\ref{fig:Stars}).
  The difference between these methods lies in the non-commutativity between
  the lensing deformations and the blurring of each star.
  A minor shortcoming of our method arises at Einstein rings
  (discussed in section~\ref{sec:AnalyticSpacetimes}),
  where the magnification diverges.
  There a star could in principle
  (though with very low probability)
  appear as an extended object,
  but in our treatment it would remain point-like.
On the other hand, blurring
first and then lensing is almost guaranteed to produce unphysical extended
streaks at the Einstein ring.

\section{Results}
\label{sec:results}

Before applying our lensing code to binary black hole systems,
we generate images of simpler analytic spacetimes. These serve both to provide
checks that our images are consistent with earlier work, and also to illustrate
general features of lensing around black holes that will appear again
in BBH images.
We then proceed to show two different configurations of BBH mergers.

To help visualize the lensing, we divide
our light source at infinity into colored
quadrants with a superimposed grid.
An external view of this sphere is shown in figure~\ref{fig:CutSphere}.
In addition to the colored sections, our light source has a bright
reference spot in the direction towards which we point our camera.
This spot will prove useful in illustrating an important feature of black hole
lensing called an Einstein ring.

\subsection{Analytic spacetimes}
\label{sec:AnalyticSpacetimes}

\begin{figure}
\centering
  \includegraphics[width=0.9\columnwidth]{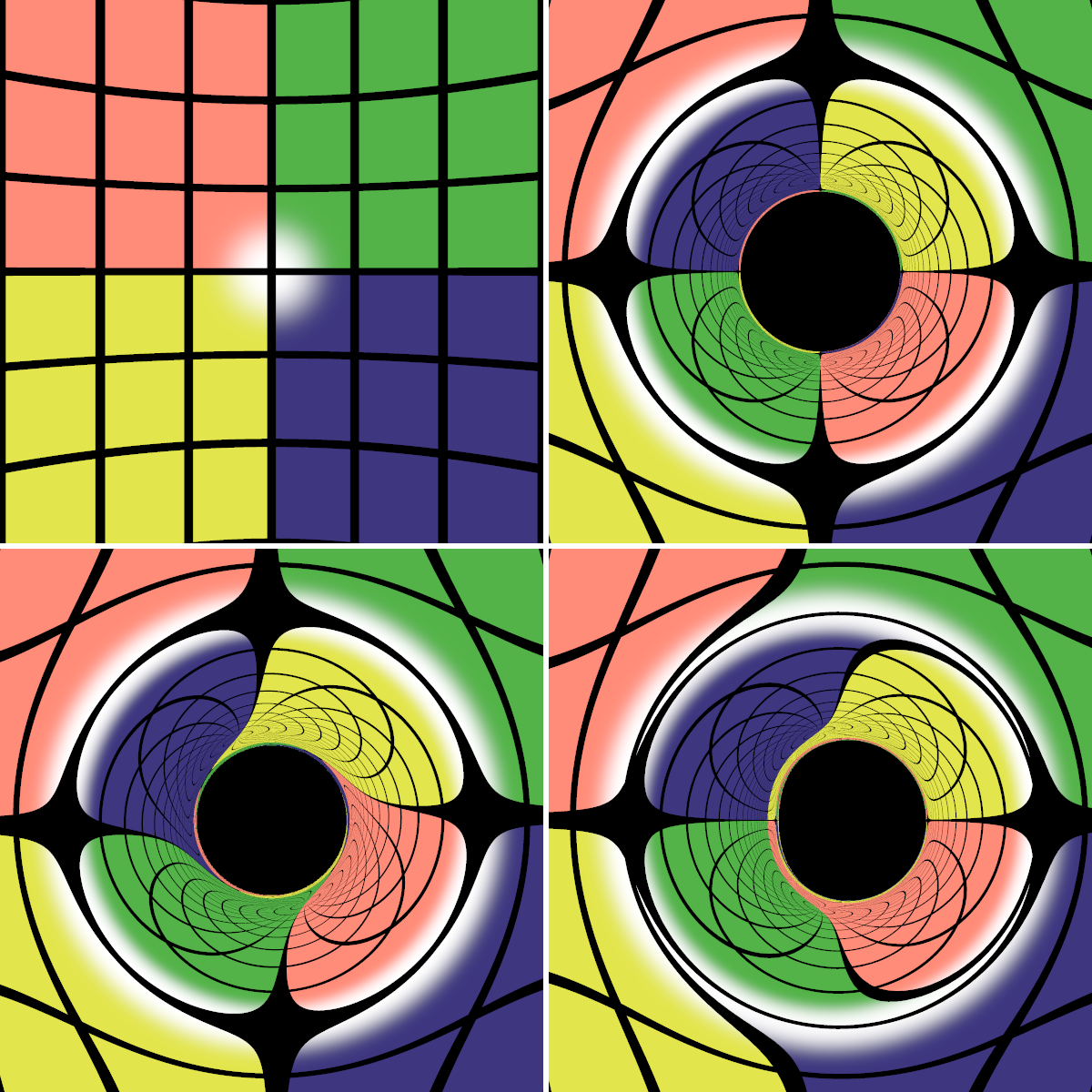}
  \caption{
    Lensing caused by various analytic spacetimes.
    For all panels, we
    use figure~\ref{fig:CutSphere} as a background,
    oriented such that the camera is pointed at the white reference dot.
    The camera has a $60 ^{\circ}$ field of view
    and is at a distance of
    $15$ Schwarzschild radii from the origin
    measured using Kerr-Schild coordinates~\cite{MTW}.
    The top row shows Minkowski and Schwarzschild spacetimes.
    The bottom row shows two views of the Kerr spacetime, with dimensionless
    spin $\chi = 0.95$,
    viewed with the camera pointing parallel to the spin axis
    of the black hole (bottom left) and
    perpendicular to the spin axis (bottom right).
  }
    \label{fig:GridMontage}
\end{figure}

In figure~\ref{fig:GridMontage}, we compare a flat space image with
the images obtained by lensing
our light source through Schwarzschild and Kerr black hole
spacetimes.
The top row from left to right shows
flat Minkowski space and a Schwarzschild black hole.
These spacetimes
are spherically symmetric, so viewing them
from different
angles produces the same lensing effects.
The bottom row shows
a Kerr black hole,
where in the left frame the spin vector
is pointing out of the page
and in the right frame it is pointing up.
Here the spin breaks the spherical
symmetry of the spacetime,
leading to different lensing effects
from different viewing
directions.

In Minkowski space in the top left image we expect no deflection of light,
which is what we observe.
The camera sees an upright image of the portion of the grid near the white
dot.
The bowing of the grid lines is an expected geometric effect of viewing a
latitude-longitude grid.

In the top right image, we see the lensing effects of a non-spinning black hole.
The black circle in the center of the image is called the shadow of the black
hole, where the hole prevents any light from reaching the camera.
Alternatively, a shadow is a region of the image where
geodesics are traced backwards in time from the camera to a black
hole.
Another
prominent feature
is that the white dot on our grid at infinity
has been lensed into a large ring,
called an Einstein ring~\cite{Einstein1936}.
Light from the point situated directly on the
opposite side of the
black hole, the antipodal point, will by symmetry be
lensed into a ring
around the black hole as observed by our camera.
Regions inside the Einstein ring correspond to photons that are deflected
by larger angles than are the Einstein ring photons; this
results in an inverted image of the reference grid inside the
Einstein ring.
A second Einstein ring can be seen near the shadow,
corresponding to light from a source behind the camera wrapping around the hole
on its way to the camera.
In fact, photons can wind an arbitrarily large number of times around the
black hole, resulting in an infinite number of Einstein rings.

The bottom row of figure~\ref{fig:GridMontage}
shows a single black hole with a large dimensionless spin
of $\chi = 0.95$.
As in the Schwarzschild case, there is an Einstein ring around
the black hole shadow
as well as image inversion inside the Einstein ring.
However, for the case of a Kerr spacetime,
the light coming from the Einstein ring does not
originate from a single point directly behind the
black hole,
but from a small
region (unless the camera is pointing directly along the spin axis).
The spin of the black hole
causes frame
dragging, where
space is dragged in the direction of the
rotation~\cite{Thirring1918, Thirring1921}.
In the bottom left image,
the spin axis of the black hole is pointing out of
the page, so space is dragged in a counterclockwise motion.
The effect of the frame dragging on the photon trajectories produces
an image in which the grid itself appears to be dragged by the spin, as is
evident when compared to the non-spinning
black hole in the top right image.
The strength of frame dragging increases closer to the black hole, which can
also be inferred from the
deformation of the background grid.

Frame dragging manifests differently in the bottom right image, where the spin
axis is pointing up.
The direction of frame dragging is out of the page on the left of the shadow
of the black hole and into the page on the right.
A photon traveling in the direction of
the frame dragging can orbit closer to the black hole
without being captured
than a photon traveling opposite
the frame dragging direction, resulting in an
asymmetrical shadow about the spin axis.
This
causes the shadow to appear offset relative to the shadow
of a Schwarzschild hole.

\subsection{Binary black hole spacetimes}
\label{sec:binary-black-hole}

\begin{figure}
\centering
%\adjincludegraphics[trim={0.09\width} {0.09\height} {0.09\width} {0.09\height},%
%    clip, width=.7\columnwidth]{\imageprefix EMtop.png}
\adjincludegraphics[trim={0.09\width} {0.09\height} {0.09\width} {0.09\height},%
    clip, width=.7\columnwidth]{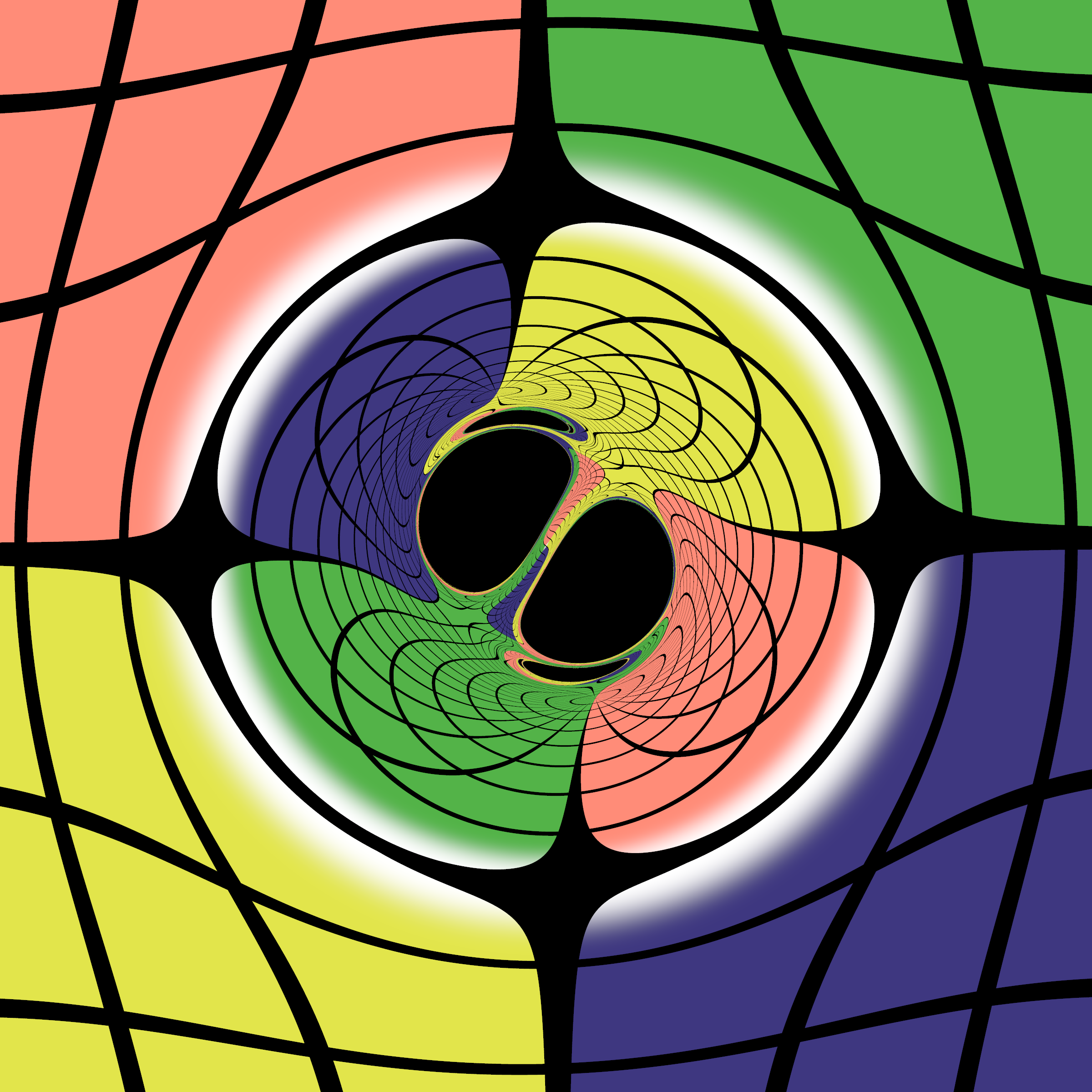}
  \caption{
    A BBH system of equal-mass black holes with no spin, viewed
    near merger with
    the orbital angular momentum out of the page.
 }
  \label{fig:EqualMassTopZoomedOut}
\end{figure}

Astrophysical black hole binaries are expected to radiate energy via
gravitational waves, leading to a long inspiral followed by
a merger, and
finally a ringdown to a steady-state single black hole.
Lensing by a final, steady-state black
hole
will look like the single black
holes already seen in figure~\ref{fig:GridMontage}.
However, the situation becomes more interesting when viewing these systems
before merger.
The first images we will present show an equal-mass
BBH with non-spinning
black holes---one of the simplest
binary inspiral spacetimes to analyze---%
shortly before merger.
The simulation we use is case~1 of
Taylor \textit{et al.}~\cite{Taylor:2013zia}.

Figure~\ref{fig:EqualMassTopZoomedOut} shows
the image of our reference grid in the presence of
this BBH, where the camera is situated such that the orbital angular
momentum is pointing out of the page.
This image bears a striking resemblance to the bottom left frame of
figure~\ref{fig:GridMontage}, excluding the details near the shadows.
This shows that, away from the shadows,
the spacetime looks
similar to a single rotating black hole,
where the lensing is dominated by the mass monopole with corrections
caused by
the angular momentum of the system.
In the single-hole case, the spin is responsible for frame dragging, whereas
here the orbital angular momentum
is responsible.

\begin{figure}
\centering
  \begin{tikzpicture}[white]
%  \node at (0,0) {\adjincludegraphics[trim={\TrimWidthEMTop\width} {\TrimWidthEMTop\height}
%  {\TrimWidthEMTop\width} {\TrimWidthEMTop\height},
%  clip,width=.7\columnwidth]{\imageprefix EMtop.png}};
  \node at (0,0) {\adjincludegraphics[trim={\TrimWidthEMTop\width} {\TrimWidthEMTop\height}
  {\TrimWidthEMTop\width} {\TrimWidthEMTop\height},
  clip,width=.7\columnwidth]{figure5.png}};

  % The coordinates below result from rescaling the zoom region (with respect to
  % the uncropped image) to the cropped image, which is centered at 0,0 and has
  % a width of .7\columnwidth
  % The formula is ((x-\TrimWidthEMTop)/(1-2\TrimWidthEMTop)-0.5)*0.7 \columnwidth
  %x's are:
  %.51, 0.535, 0.595, 0.62
  \pgfmathsetlengthmacro{\xl}{0.01591\columnwidth};
  \pgfmathsetlengthmacro{\xr}{0.05568\columnwidth};
  \pgfmathsetlengthmacro{\yb}{0.15114\columnwidth};
  \pgfmathsetlengthmacro{\yt}{0.19091\columnwidth};
  \begin{scope}
    \begin{pgfinterruptboundingbox}
    \clip (\xl,\yt) rectangle (\xr,\yb) (-100,-100) rectangle (100,100);
    \end{pgfinterruptboundingbox}
    \draw[line width=2pt] (\xl,\yb) rectangle (\xr,\yt);
  \end{scope}
%  \node[draw,ultra thick,inner sep=0] at (3,1.5)
%  (box){\includegraphics[width=.3\columnwidth]{\imageprefix EMtopZoom.png}};
  \node[draw,ultra thick,inner sep=0] at (3,1.5)
  (box){\includegraphics[width=.3\columnwidth]{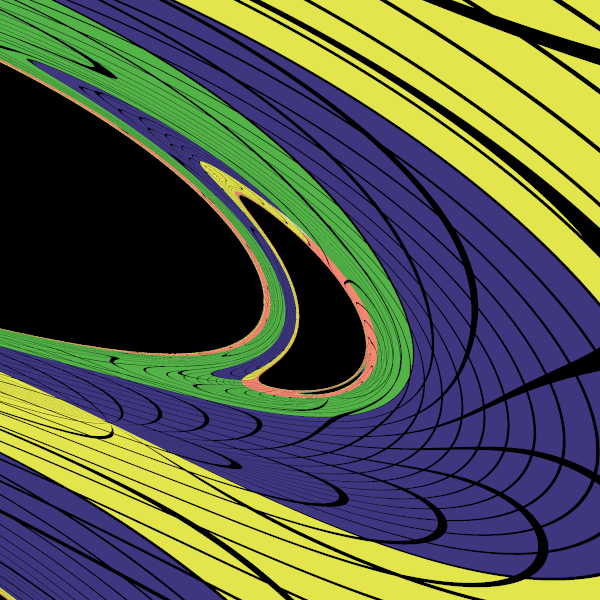}};
  \end{tikzpicture}
  \caption{
    A cropped version of figure~\ref{fig:EqualMassTopZoomedOut} in order to
    show more detail near the black hole shadows.
    A small portion of the
    image (outlined) is enlarged and inset, where a smaller
    eyebrow is clearly visible.
  }
  \label{fig:EqualMassTop}
\end{figure}
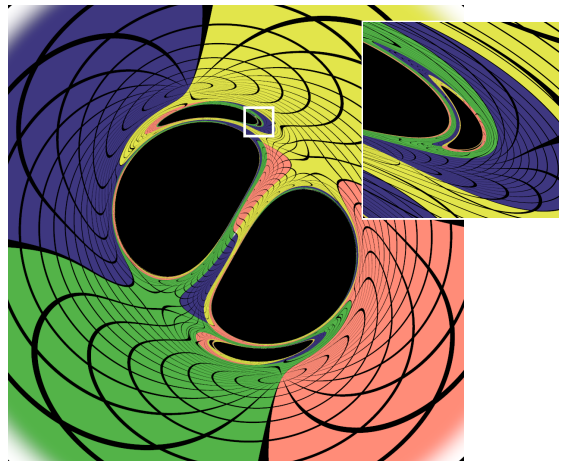

Focusing on the inner portion of the image,
we observe that the binary lensing is
markedly different from the Schwarzschild or Kerr cases.
Figure~\ref{fig:EqualMassTop} shows a cropped version of
figure~\ref{fig:EqualMassTopZoomedOut},
emphasizing the structure of the shadows.
As might be expected,
there are two prominent shadows visible, each associated
with one of the two black holes.
We also see a narrow secondary shadow (an ``eyebrow''~\cite{Yumoto2012})
close to the outside of each
primary shadow.
These secondary shadows correspond to one black hole (BH)
casting a shadow which is lensed by the other BH on the way to the camera.
Equivalently, they are image regions where geodesics are traced
backwards from the camera to a BH,
but bend around at least one BH on the way there.
The first pair of eyebrows is evident in figure~\ref{fig:EqualMassTop};
however, we can resolve a pair of smaller eyebrows,
shown in the inset.

\begin{figure}
\centering
  \begin{tikzpicture}[white]
%  \node at (0,0) {\adjincludegraphics[trim={\TrimWidth\width} {\TrimWidth\height}
%  {\TrimWidth\width} {\TrimWidth\height},
%  clip,width=.7\columnwidth]{\imageprefix EMside.png}};
  \node at (0,0) {\adjincludegraphics[trim={\TrimWidth\width} {\TrimWidth\height}
  {\TrimWidth\width} {\TrimWidth\height},
  clip,width=.7\columnwidth]{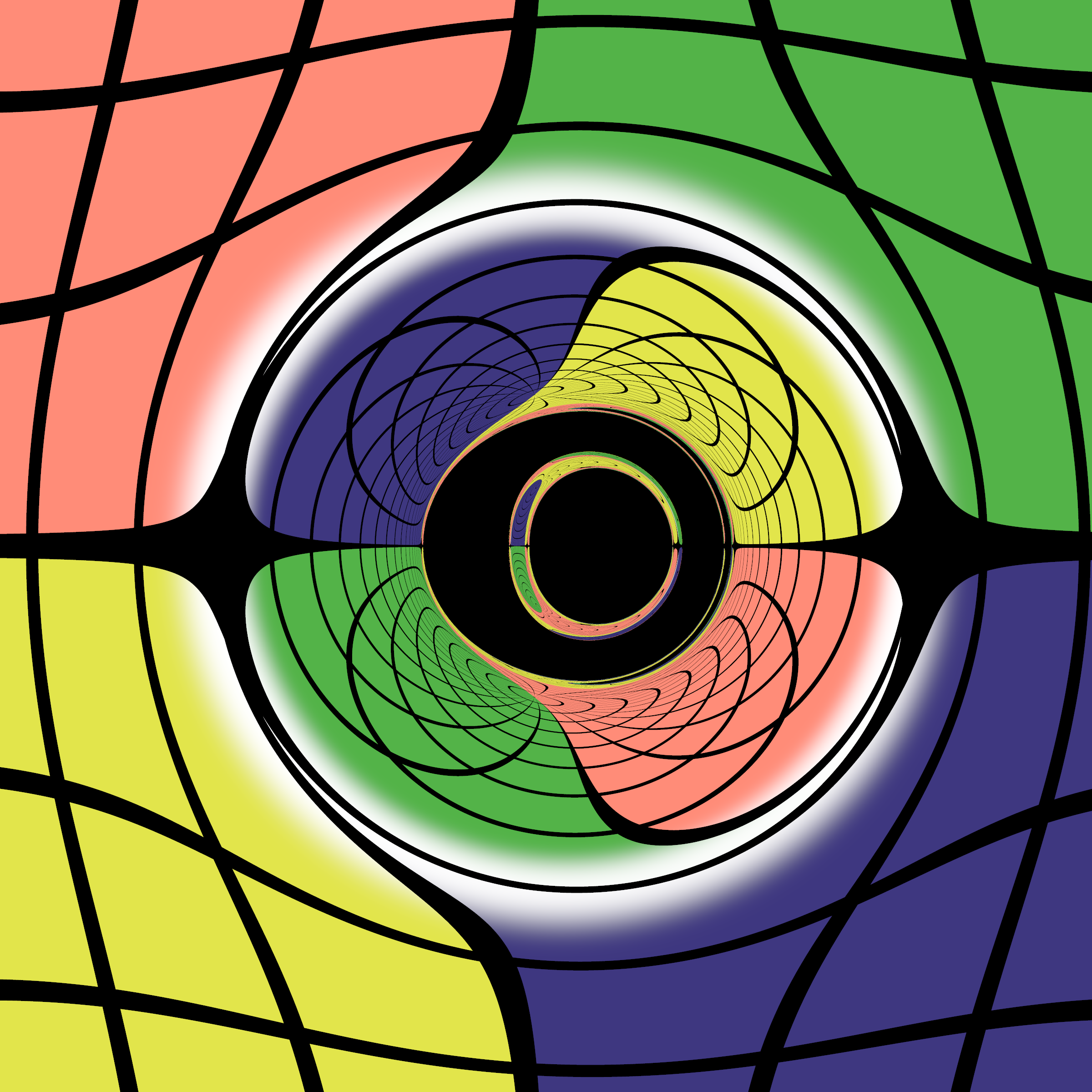}};

  % The coordinates below result from rescaling the zoom region (with respect to
  % the uncropped image) to the cropped image, which is centered at 0,0 and has
  % a width of .7\columnwidth
  % The formula is ((x-\TrimWidth)/(1-2\TrimWidth)-0.5)*0.7 \columnwidth
  % where x is given via
  % 0.65+0.03*0.3667, 0.65+0.03*0.7, 0.485+0.03*0.3333, or 0.485+0.03*0.6667
  \pgfmathsetlengthmacro{\xl}{0.20871\columnwidth};
  \pgfmathsetlengthmacro{\xr}{0.22167\columnwidth};
  \pgfmathsetlengthmacro{\yb}{-0.00648\columnwidth};
  \pgfmathsetlengthmacro{\yt}{+0.00648\columnwidth};
  \begin{scope}
    \begin{pgfinterruptboundingbox}
    \clip (\xl,\yt) rectangle (\xr,\yb) (-100,-100) rectangle (100,100);
    \end{pgfinterruptboundingbox}
    \draw[line width=2pt] (\xl,\yb) rectangle (\xr,\yt);
  \end{scope}
%  \node[draw,ultra thick,inner sep=0] at (-3,1.5)
%  (box){\includegraphics[width=.3\columnwidth]{\imageprefix EMsideZoom.png}};
  \node[draw,ultra thick,inner sep=0] at (-3,1.5)
  (box){\includegraphics[width=.3\columnwidth]{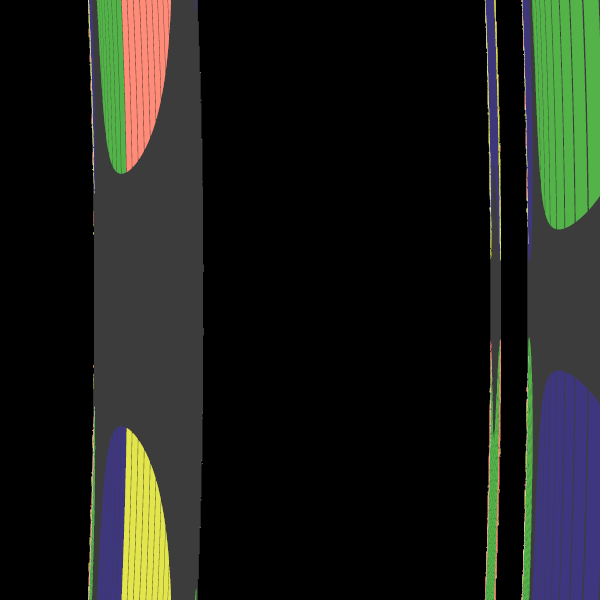}};

  % Uncomment the following and adjust to try and center large image.
  %\path (box.west) -- (5,0);
  \end{tikzpicture}
  \caption{
    The same system as figure~\ref{fig:EqualMassTop}, viewed such that
    the orbital angular momentum of the system is pointing up.
    Note that the grid lines in the inset are shown in gray here to distinguish
    them from the black hole shadows.
  }
  \label{fig:EqualMassSide}
\end{figure}

We show another view of the same system
in figure~\ref{fig:EqualMassSide}.
Here the camera is looking at the system edge on,
such that the orbital angular momentum
is pointing up.
We see again an overall similarity with the corresponding orientation
of Kerr spacetime (the bottom-right frame of figure~\ref{fig:GridMontage}),
indicating the dominant effects of the mass
and angular momentum in these images.
We can see a primary shadow for each black hole,
but in this configuration one black hole is located roughly behind
the other and as a result its shadow gets lensed into a dark ring.
Extending along the right side of this ring we see a long
thin eyebrow,
which is shown in the inset,
along with another, smaller, eyebrow.

\begin{figure}
\centering
  \includegraphics[width=0.7\columnwidth]{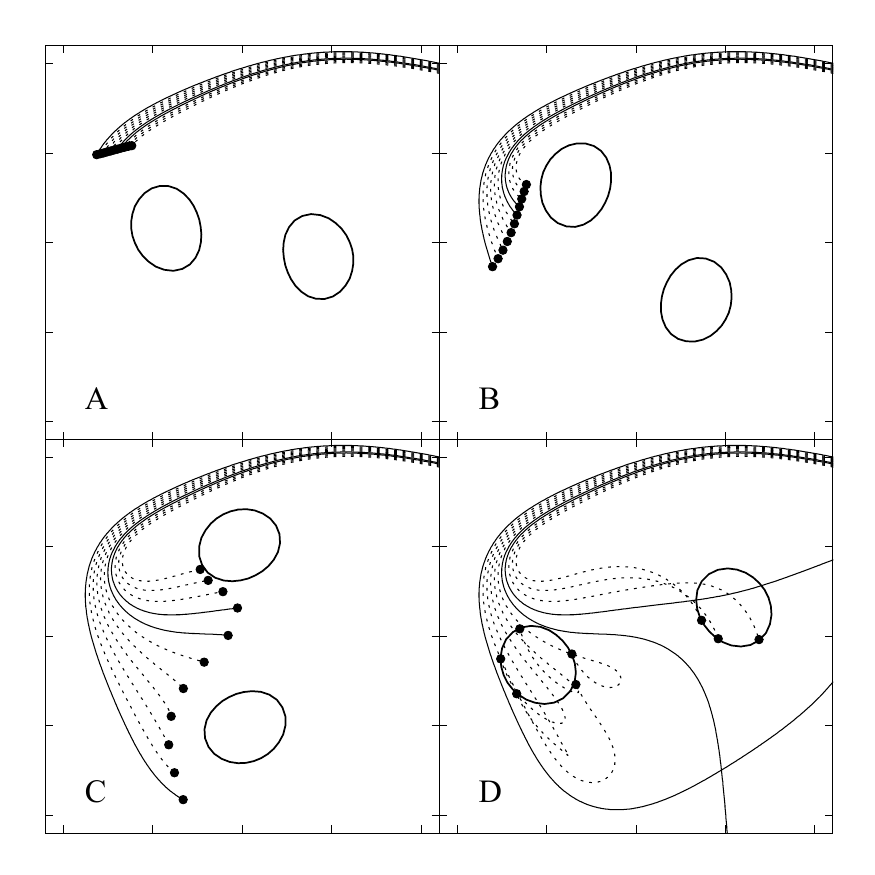}
  \caption{
    Geodesic trajectories plotted in relation to the black hole event horizons
    during the lensing evolution for figure~\ref{fig:EqualMassSide}.
    Each frame shows a snapshot in time, with the
    dots representing the current positions of the geodesics, and the lines
    indicating the trajectories from the camera.
    The solid and dashed lines indicate whether the geodesics originate from
    infinity or from a black hole, respectively.
  }
  \label{fig:TrajectoryFigure}
\end{figure}

To illustrate how photon trajectories behave near shadows,
we plot trajectories of a few geodesics on the
horizontal line passing through the middle of figure~\ref{fig:EqualMassSide}
near the eyebrow.
Figure~\ref{fig:TrajectoryFigure} shows four snapshots
of these
trajectories in time,
with
their current locations in each frame denoted by large dots.
It is easiest to consider these trajectories as
we evolve them,
out of the camera and backwards in time,
to see where they came from.
In frames A--C, we see the trajectories under consideration start
close together then diverge significantly, demonstrating how
nearby pixels on the image can correspond to
vastly different physical locations.
In frame~D we see the entire trajectories. A few extend to infinity, but most
terminate on the black holes; these are denoted by solid lines and dotted
lines, respectively. Only the trajectories extending to infinity
result in a photon reaching the camera;
those that reach the hole on the right of frame~D correspond to the
primary ring-like shadow in figure~\ref{fig:EqualMassSide},
while those that reach the left hole correspond to the larger eyebrow
visible on the right side of figure~\ref{fig:EqualMassSide}.
Note that the black holes are orbiting rapidly,
so they move significantly while the photons pass through
the system.

{
  \begin{figure}
  %Global coordinates are simply camera coordinates of the desired plot range
%Min and Max with no prefix correspond to the min and max of the dataset,
% which is really refinement coordinates
%Simul coordinates are the min and max coordinates passed to Camera.input
% to truncate the width of the simulation
\pgfmathsetmacro\MinPlotZero{\TrimWidth}
\pgfmathsetmacro\MaxPlotZero{1 - \MinPlotZero}
\pgfmathsetmacro\XDiffPlotZero{\MaxPlotZero - \MinPlotZero}
%Only in plot zero, the global coordinates are equal to the min and max
% for the picture.  This is because plot zero uses the full camera plane
\pgfmathsetmacro\GlobalMinPlotZero{\MinPlotZero}
\pgfmathsetmacro\GlobalMaxPlotZero{\MaxPlotZero}
\pgfmathsetmacro\GlobalDiffPlotZero{\GlobalMaxPlotZero - \GlobalMinPlotZero}

%General locations and properties
\pgfmathsetmacro\Width{0.7}
\pgfmathsetmacro\Height{0.12}
\pgfmathsetmacro\Spacing{0.065}
\pgfmathsetmacro\VerticalSegment{0.015}
\pgfmathsetmacro\TopPlotZero{3*\Height + 2*\Spacing}
\pgfmathsetmacro\BottomPlotZero{2*(\Height + \Spacing)}
\pgfmathsetmacro\BottomPlotZeroMinusSeg{\BottomPlotZero - \VerticalSegment}
\pgfmathsetmacro\TopPlotOne{\BottomPlotZero - \Spacing}
\pgfmathsetmacro\BottomPlotOne{\Height + \Spacing}
\pgfmathsetmacro\BottomPlotOneMinusSeg{\BottomPlotOne - \VerticalSegment}
\pgfmathsetmacro\TopPlotTwo{\Height}

\begin{tikzpicture}
\begin{axis} [
  name=plot0,
  at={(0,\BottomPlotZero\columnwidth)},
  scale only axis,
  width=\Width\columnwidth,
  height=\Height\columnwidth,
  xmin=\MinPlotZero,xmax=\MaxPlotZero,
  xtick=\empty,
  ytick={1,2,3},
  yticklabels={BH 1, $\infty$, BH 2},
  ytick pos=left,
]
  \addplot[no marks] plot
    file {SurfaceIDsZoom0.dat};
\end{axis}

%draw the connecting lines
%These are the mins and maxs for the simulation itself
\pgfmathsetmacro\SimulMinPlotOne{0.65}
\pgfmathsetmacro\SimulMaxPlotOne{0.68}
%These are the desired mins and maxs relative to the plot0 coordinates
\pgfmathsetmacro\GlobalMinPlotOne{0.661}
\pgfmathsetmacro\GlobalMaxPlotOne{0.671}
\pgfmathsetmacro\GlobalDiffPlotOne{\GlobalMaxPlotOne - \GlobalMinPlotOne}
\pgfmathsetmacro\ZoomMinPlotZero{\Width*((\GlobalMinPlotOne-\GlobalMinPlotZero)/(\GlobalDiffPlotZero)};
\pgfmathsetmacro\ZoomMaxPlotZero{\Width*((\GlobalMaxPlotOne-\GlobalMinPlotZero)/(\GlobalDiffPlotZero)};

%  \draw[dotted] (\ZoomMinPlotZero\columnwidth, \BottomPlotZero\columnwidth)
%    -- (\ZoomMinPlotZero\columnwidth, \TopPlotZero\columnwidth);
%  \draw[dotted] (\ZoomMaxPlotZero\columnwidth, \BottomPlotZero\columnwidth)
%    -- (\ZoomMaxPlotZero\columnwidth, \TopPlotZero\columnwidth);
\draw[thick] (\ZoomMinPlotZero\columnwidth, \BottomPlotZero\columnwidth)
  -- (\ZoomMinPlotZero\columnwidth, \BottomPlotZeroMinusSeg\columnwidth);
\draw[thick] (\ZoomMaxPlotZero\columnwidth, \BottomPlotZero\columnwidth)
  -- (\ZoomMaxPlotZero\columnwidth, \BottomPlotZeroMinusSeg\columnwidth);
\draw[dashed, thick] (\ZoomMinPlotZero\columnwidth, \BottomPlotZeroMinusSeg\columnwidth) --
  (0, \TopPlotOne\columnwidth);
\draw[dashed, thick] (\ZoomMaxPlotZero\columnwidth, \BottomPlotZeroMinusSeg\columnwidth) --
  (\Width\columnwidth, \TopPlotOne\columnwidth);

%This data should match exactly with a zoombox between
% x = 0.655 and 0.675
% y = 0.49 and 0.51
% relative to the full run
% however, the run outputs coordinates between 0 and 1, even though
% I requested x between 0.65 and 0.68
\pgfmathsetmacro\SimulDiffPlotOne{\SimulMaxPlotOne - \SimulMinPlotOne}
\pgfmathsetmacro\MinPlotOne{
  (\GlobalMinPlotOne - \SimulMinPlotOne)/\SimulDiffPlotOne}
\pgfmathsetmacro\MaxPlotOne{
  (\GlobalMaxPlotOne - \SimulMinPlotOne)/\SimulDiffPlotOne}
\pgfmathsetmacro\XDiffPlotOne{\MaxPlotOne - \MinPlotOne}

\begin{axis} [
  name=plot1,
  at={(0,\BottomPlotOne\columnwidth)},
  scale only axis,
  height=\Height\columnwidth,
  width=\Width\columnwidth,
  xmin=\MinPlotOne,xmax=\MaxPlotOne,
  xtick=\empty,
  ytick={1,2,3},
  yticklabels={BH 1, $\infty$, BH 2},
  ytick pos=left,
]
  \addplot[no marks] plot
    file {SurfaceIDsZoom1.dat};
\end{axis}

%draw the connecting lines
%These are the mins and maxs for the simulation itself
\pgfmathsetmacro\SimulMinPlotTwo{0.66965}
\pgfmathsetmacro\SimulMaxPlotTwo{0.66992}
\pgfmathsetmacro\SimulDiffPlotTwo{\SimulMaxPlotTwo - \SimulMinPlotTwo}
%These are arbitrarily chosen
\pgfmathsetmacro\MinPlotTwo{0.32}
\pgfmathsetmacro\MaxPlotTwo{0.53}
%These are the desired mins and maxs relative to the plot1 coordinates
\pgfmathsetmacro\GlobalMinPlotTwo{\MinPlotTwo*\SimulDiffPlotTwo + \SimulMinPlotTwo}
\pgfmathsetmacro\GlobalMaxPlotTwo{\MaxPlotTwo*\SimulDiffPlotTwo + \SimulMinPlotTwo}

\pgfmathsetmacro\ZoomMinPlotOne{\Width*(\GlobalMinPlotTwo-\GlobalMinPlotOne)/(\GlobalDiffPlotOne)};
\pgfmathsetmacro\ZoomMaxPlotOne{\Width*(\GlobalMaxPlotTwo-\GlobalMinPlotOne)/(\GlobalDiffPlotOne)};

%  \draw[dotted] (\ZoomMinPlotOne\columnwidth, \BottomPlotOne\columnwidth)
%    -- (\ZoomMinPlotOne\columnwidth, \TopPlotOne\columnwidth);
%  \draw[dotted] (\ZoomMaxPlotOne\columnwidth, \BottomPlotOne\columnwidth)
%    -- (\ZoomMaxPlotOne\columnwidth, \TopPlotOne\columnwidth);
\draw[thick] (\ZoomMinPlotOne\columnwidth, \BottomPlotOne\columnwidth)
  -- (\ZoomMinPlotOne\columnwidth, \BottomPlotOneMinusSeg\columnwidth);
\draw[thick] (\ZoomMaxPlotOne\columnwidth, \BottomPlotOne\columnwidth)
  -- (\ZoomMaxPlotOne\columnwidth, \BottomPlotOneMinusSeg\columnwidth);
\draw[dashed, thick] (\ZoomMinPlotOne\columnwidth, \BottomPlotOneMinusSeg\columnwidth)
  -- (0, \TopPlotTwo\columnwidth);
\draw[dashed, thick] (\ZoomMaxPlotOne\columnwidth, \BottomPlotOneMinusSeg\columnwidth)
  -- (\Width\columnwidth, \TopPlotTwo\columnwidth);
%\draw[ultra thick] (\ZoomMinPlotOne\columnwidth, \BottomPlotOne\columnwidth)
%  -- (\ZoomMaxPlotOne\columnwidth, \BottomPlotOne\columnwidth);

\begin{axis} [
  name=plot2,
  at={(0, 0)},
  scale only axis,
  height=\Height\columnwidth,
  width=\Width\columnwidth,
  xmin=\MinPlotTwo,xmax=\MaxPlotTwo,
  xtick=\empty,
  ytick={1,2,3},
  yticklabels={BH 1, $\infty$, BH 2},
  ytick pos=left,
]
  %\addplot {sin(x)};
  \addplot[no marks] plot
    file {SurfaceIDsZoom2.dat};
\end{axis}
\end{tikzpicture}
  \caption{
    Plots identifying the origins of photons along the
    horizontal line through the center of figure~\ref{fig:EqualMassSide}.
    Photons coming from infinity are labeled $\infty$, and the shadows
    are labeled either BH~1 or BH~2.
    The first plot corresponds to the main
    portion of figure~\ref{fig:EqualMassSide}.
    The second plot focuses on the zoomed square in the inset
      of
    figure~\ref{fig:EqualMassSide}, showing a small feature of the first plot.
    The third plot zooms to a similar feature of the second plot.
    This figure demonstrates a striking self-similarity of the
    lensing structure of a binary black hole system.
  }
  \label{fig:SelfSimilarity}
  \end{figure}
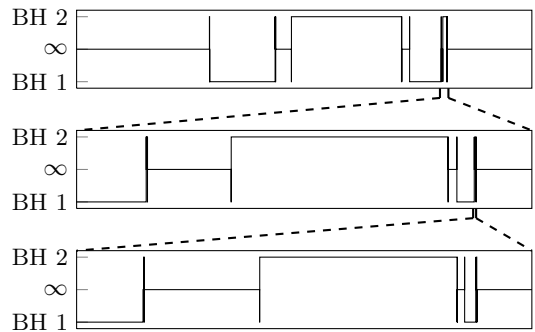
}

We can also uniquely identify which black hole
casts each shadow, which enables us to show in
figure~{\ref{fig:SelfSimilarity}}
the origin of the photons along
the horizontal line across the center of
figure~{\ref{fig:EqualMassSide}}.
We arbitrarily
label the large shadow in the middle of
figure~\ref{fig:EqualMassSide} as BH~2, and the
ring-like shadow as BH~1.
Regions where photons reach the camera from infinity are labeled $\infty$.
The top plot in figure~\ref{fig:SelfSimilarity} shows the origin of
    the photons that reach the camera along the entire middle horizontal
    line in figure~{\ref{fig:EqualMassSide}}. We
see that each transition from $\infty$ to either of the BHs
includes transitions to the other BH.
Even though we cannot resolve them numerically, each vertical line in principle
contains infinitely many transitions.
To illustrate this idea,
the second plot in figure~\ref{fig:SelfSimilarity}
investigates the group of shadows indicated by the zoomed
inset of figure~\ref{fig:EqualMassSide}.
Here we find a structure which resembles the first plot.
The third plot in figure~\ref{fig:SelfSimilarity}
zooms to a similar group of shadows on the right side of the
second plot to again reveal the same structure.
This figure clearly shows evidence of self-similarity in the structure of
BBH lensing, where the smaller length scales explore more photon orbits through
the system.
Furthermore, the structure of shadows in BBH lensing is more complex than
figures~\ref{fig:EqualMassTop} and~\ref{fig:EqualMassSide} appear to suggest.
The shadows these images focus on are merely some of the largest visible
shadows, associated with simpler geodesic orbits around the binary.

If we consider this equal-mass BBH earlier in the
inspiral
when its separation is large, the
black holes are only weakly interacting.
Therefore most camera viewpoints of
this binary will yield images with two primary shadows,
one for each black hole.
Each shadow will be
similar to an isolated Schwarzschild or Kerr shadow but with the addition
of small eyebrows.
However, when the binary is viewed edge-on and
the black holes are nearly aligned
with the camera, we see an interesting image.

\begin{figure}
\centering
%\adjincludegraphics[trim={0.15\width} {0.15\height} {0.15\width} {0.15\height},%
%    clip, width=.7\columnwidth]{\imageprefix EMLargeSep.png}
\adjincludegraphics[trim={0.15\width} {0.15\height} {0.15\width} {0.15\height},%
    clip, width=.7\columnwidth]{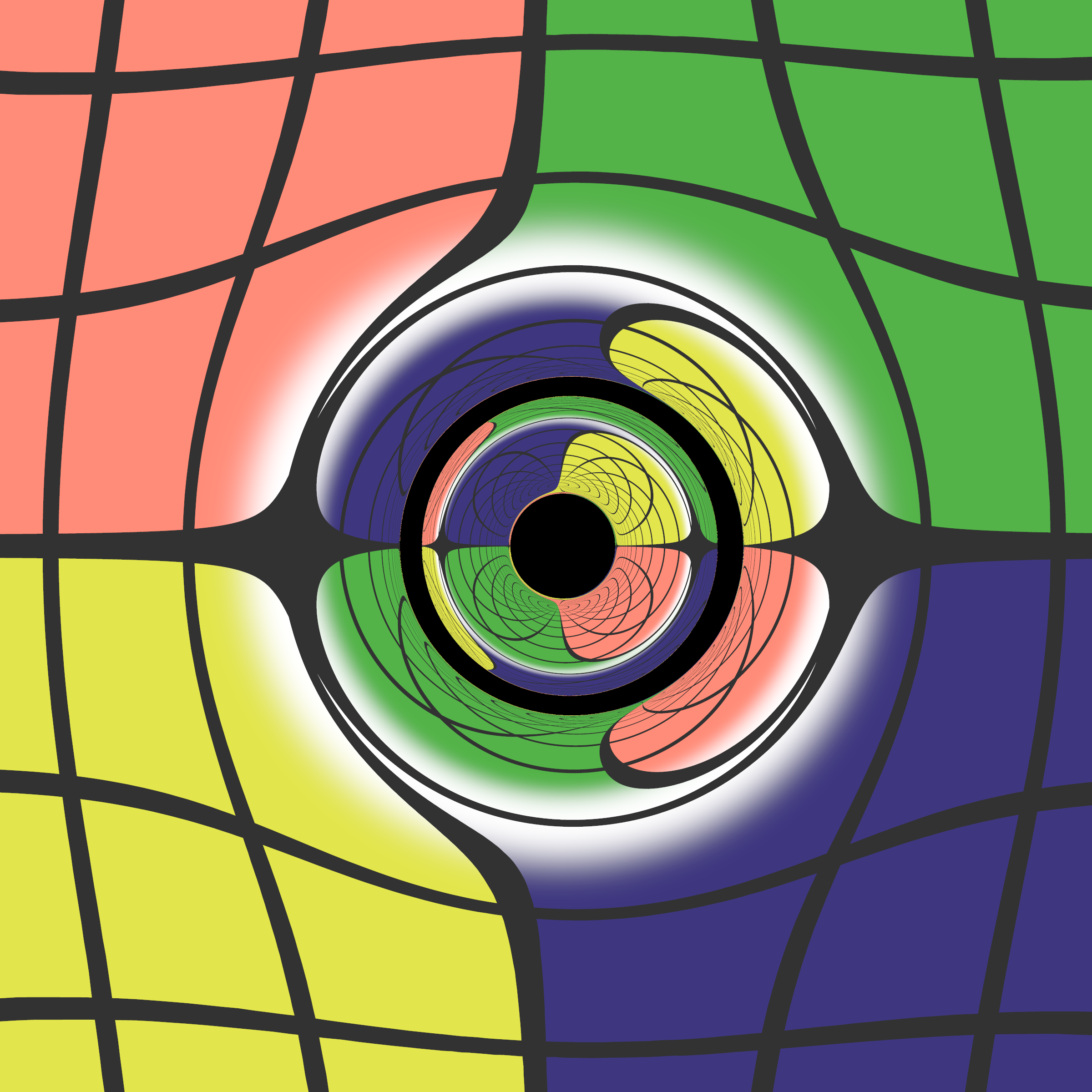}
  \caption{
    A BBH system of equal-mass black holes with no spin, viewed hundreds of
    orbits before merger, with the orbital angular momentum pointing up.
    The distance from the camera to the closer black hole in this figure
    is the same as in figure~\ref{fig:EqualMassSide}.
    Note that the grid lines are shown in gray here to distinguish
    them from the black hole shadows.
 }
  \label{fig:EqualMassLargeSep}
\end{figure}

Figure~\ref{fig:EqualMassLargeSep} shows the equal-mass binary
in this configuration,
hundreds of orbits before merger.
Just as in figure~\ref{fig:EqualMassSide}, the
more distant black hole is
lensed into a ring-like shadow; however, the ring is
thinner here, primarily
because of the large separation of the binary.
The angular momentum causes
the lensed grid outside the shadows to strongly resemble lensing by a
Kerr black
hole rather than lensing by a Schwarzschild black hole.
In addition to the usual primary Einstein ring,
another ring is visible between these
shadows.  Both of these rings correspond
to the same
source of light, which is in front of the camera and behind the BBH.
The second Einstein ring is caused by photons following
an ``S''-shaped trajectory through the system.

\begin{figure}
  \centering
  \begin{tikzpicture}[white]
%  \node at (0,0) {\adjincludegraphics[trim={\TrimWidthForGeneric\width}
%    {\TrimWidthForGeneric\height} {\TrimWidthForGeneric\width} {\TrimWidthForGeneric\height},
%  clip,width=.7\columnwidth]{\imageprefix GenericPara.png}};
  \node at (0,0) {\adjincludegraphics[trim={\TrimWidthForGeneric\width}
    {\TrimWidthForGeneric\height} {\TrimWidthForGeneric\width} {\TrimWidthForGeneric\height},
  clip,width=.7\columnwidth]{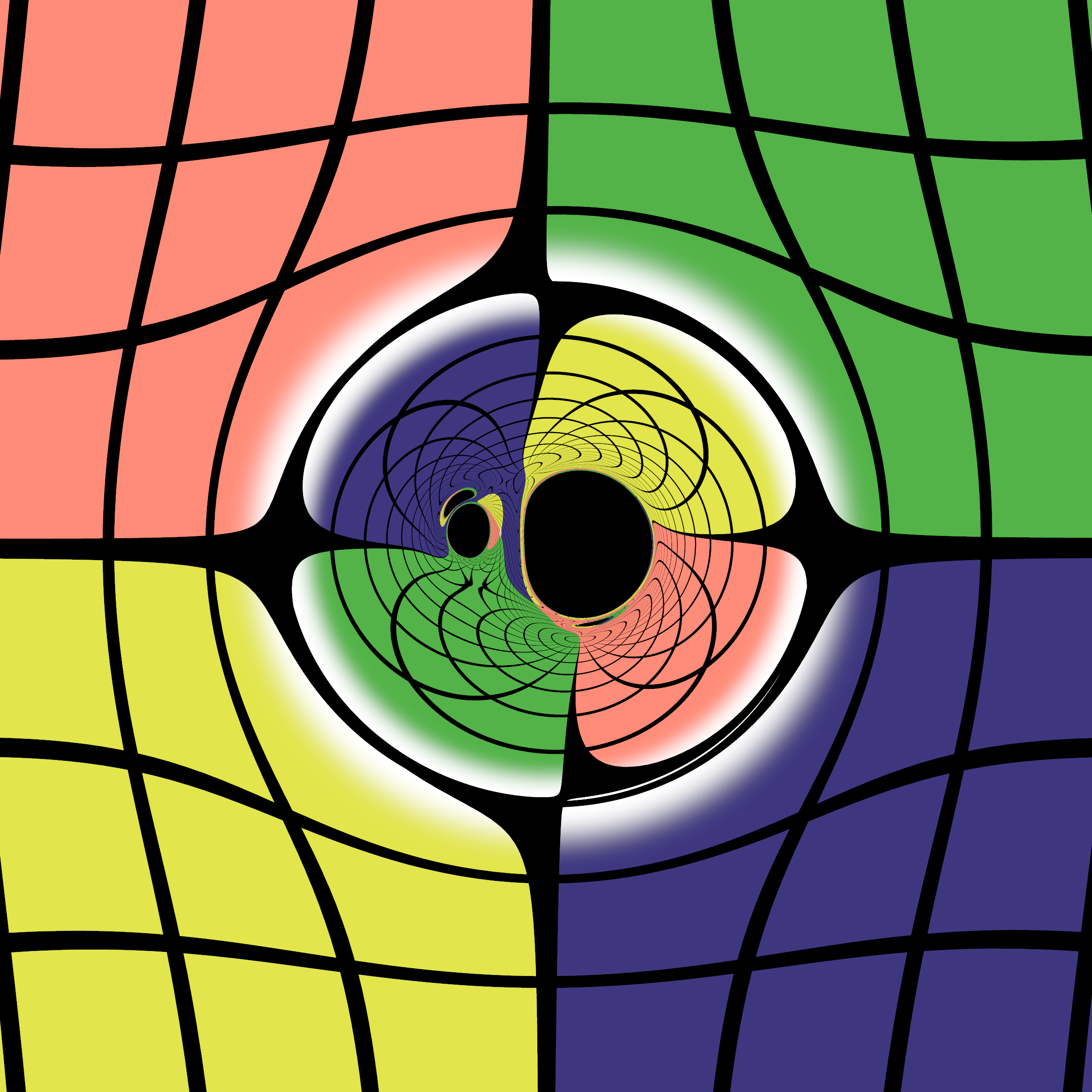}};

  % The coordinates below result from rescaling the zoom region (with respect to
  % the uncropped image) to the cropped image, which is centered at 0,0 and has
  % a width of .7\columnwidth
  % The formula is ((x-\GenericTrimWidth)/(1-2\GenericTrimWidth)-0.5)*0.7 \columnwidth
  % where x is given via
  % 0.425+0.0125*0.625, 0.425+0.0125*0.875, 0.54+0.0125*0.375, or
    % 0.54+0.0125*0.625
  %\pgfmathsetlengthmacro{\xl}{-0.11198\columnwidth};
  %\pgfmathsetlengthmacro{\xr}{-0.10677\columnwidth};
  %\pgfmathsetlengthmacro{\yb}{ 0.07448\columnwidth};
  %\pgfmathsetlengthmacro{\yt}{ 0.07969\columnwidth};

  %Note: The following are calculated to make the box 30% bigger in size,
  % meaning the 0.625, 0.875 and 0.375, 0.625 factors in the commented
  % equations have been modified accordingly
  \pgfmathsetlengthmacro{\xl}{-0.11276\columnwidth};
  \pgfmathsetlengthmacro{\xr}{-0.10600\columnwidth};
  \pgfmathsetlengthmacro{\yb}{ 0.07370\columnwidth};
  \pgfmathsetlengthmacro{\yt}{ 0.08047\columnwidth};
  \begin{scope}
    \begin{pgfinterruptboundingbox}
    \clip (\xl,\yt) rectangle (\xr,\yb) (-100,-100) rectangle (100,100);
    \end{pgfinterruptboundingbox}
    \draw[thick] (\xl,\yb) rectangle (\xr,\yt);
  \end{scope}
%  \node[draw,ultra thick,inner sep=0] at (-3,1.5)
%  (box){\includegraphics[width=.3\columnwidth]{\imageprefix
%    GenericParaZoom.png}};
  \node[draw,ultra thick,inner sep=0] at (-3,1.5)
  (box){\includegraphics[width=.3\columnwidth]{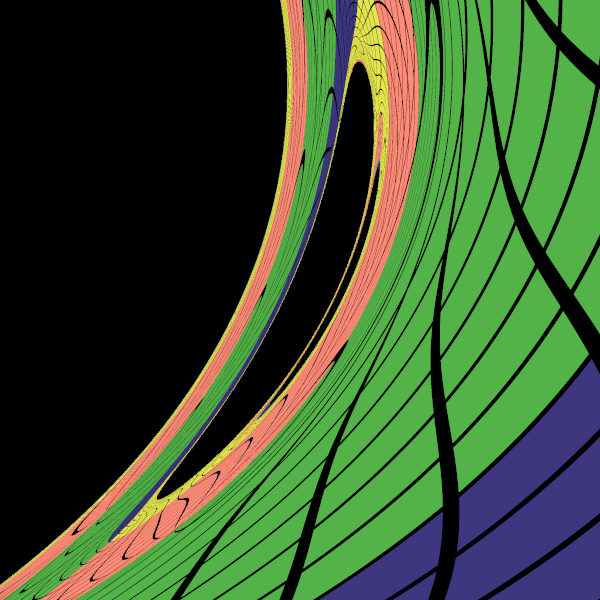}};

  % Uncomment the following and adjust to try and center large image.
  %\path (box.west) -- (5,0);
  \end{tikzpicture}
  \caption{
    A view of a binary inspiral of mass ratio
    $m_1/m_2=3$
    near merger,
    with the orbital angular
    momentum approximately pointing out of the page.
    The black hole spins are $\chi_1=0.7$ and $\chi_2=0.3$ in arbitrary
    directions.
    This figure is analogous to figure~\ref{fig:EqualMassTop}.
    As in previous figures,
    a small portion
    of the image is enlarged and inset, displaying additional
    eyebrows.
  }
  \label{fig:GenericTop}
\end{figure}

The second binary system we consider
is a fully generic black hole binary with a mass ratio of $m_1/m_2=3$ and
black hole spins of $\chi_1=0.7$ and $\chi_2=0.3$ in arbitrary directions. This
is case 4 of
Taylor \textit{et al.}~\cite{Taylor:2013zia}.
In figure~\ref{fig:GenericTop} we see a top view of this system,
in analogy with
what is presented in figure~\ref{fig:EqualMassTop}.
Away from the shadows,
the lensing is similar to a single black hole with spin,
as was seen with the equal-mass binary images.
This appears to be a generic feature of
lensing from orbiting BBHs.
We can
clearly see that the symmetry present in the equal-mass system is gone.
The unequal masses evidently change the relative sizes of not only the
primary shadows, but all additional shadows as well.
The inset in figure~\ref{fig:GenericTop} zooms to show
two successively smaller eyebrows near the small black hole's primary shadow.
However, the effects of the black holes' spins are not at all clear from this
viewpoint.

\begin{figure}
  \centering
%\adjincludegraphics[trim={\TrimWidthForGeneric\width} {\TrimWidthForGeneric\height} {\TrimWidthForGeneric\width} {\TrimWidthForGeneric\height},%
%    clip, width=.7\columnwidth]{\imageprefix GenericSide}
\adjincludegraphics[trim={\TrimWidthForGeneric\width} {\TrimWidthForGeneric\height} {\TrimWidthForGeneric\width} {\TrimWidthForGeneric\height},%
    clip, width=.7\columnwidth]{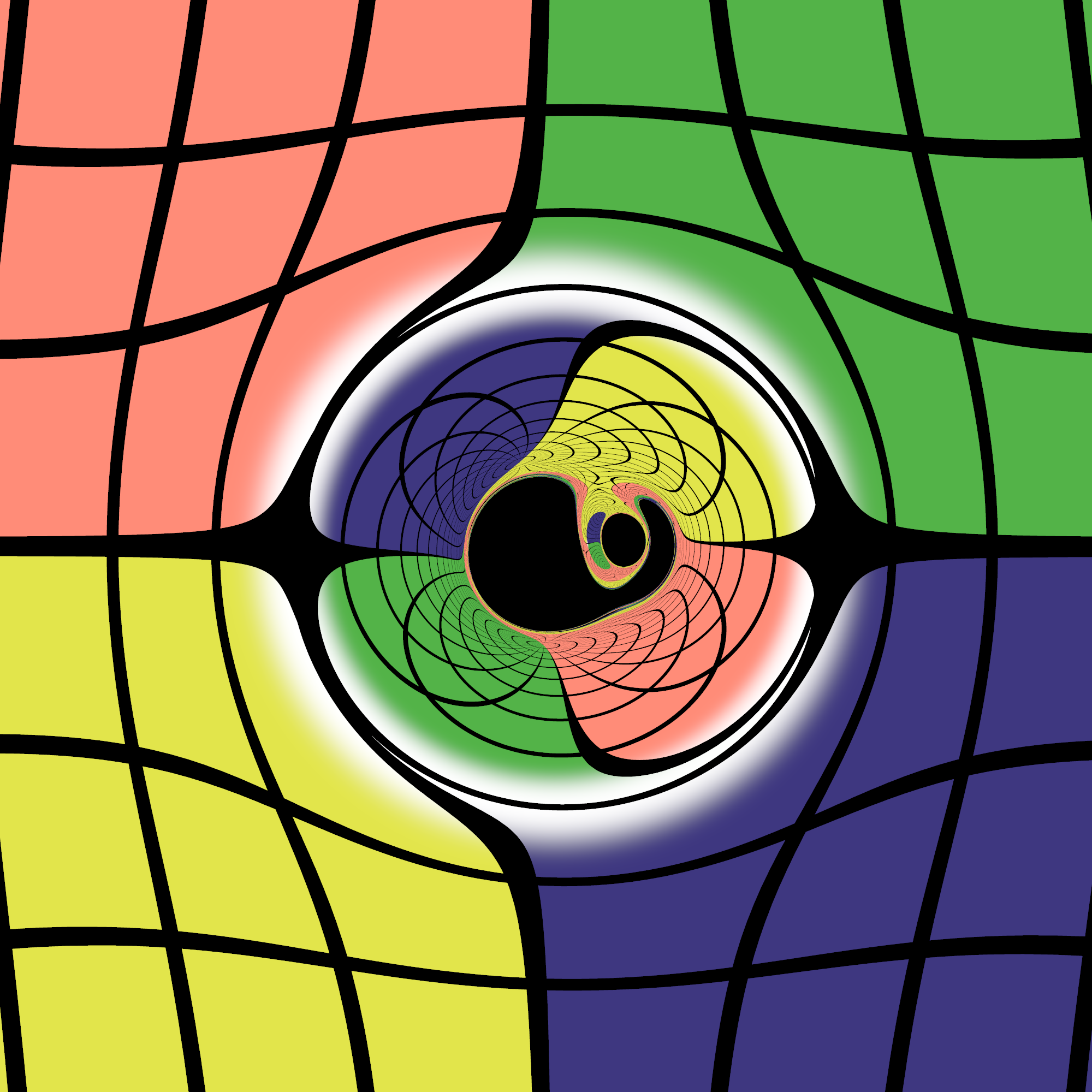}
  % current image is from GenericT6701Y0.3Z1.5A70
  \caption{
    Another view of the BBH in figure~\ref{fig:GenericTop}, but with
    orbital angular momentum pointing up.
    The camera parameters are otherwise identical.
    This figure is analogous to figure~\ref{fig:EqualMassSide}; however,
    because of the asymmetry from the black hole spins,
    the larger black hole's shadow is not lensed
    completely around the small black hole.
  }
  \label{fig:GenericSide}
\end{figure}

In figure~\ref{fig:GenericSide} we see the same binary
as in
figure~\ref{fig:GenericTop},
viewed with the orbital angular momentum pointing upward,
in analogy with figure~\ref{fig:EqualMassSide}.
We again see that, away from the shadows, the system looks like a Kerr
black hole.
The unequal mass ratio is apparent here, with the smaller black hole lensing the
shadow of the larger black hole into a partial ring.
If it were not for the black hole spins, the lensing
by the binary would be
symmetric, giving either a ring-like shadow
similar to figure~\ref{fig:EqualMassSide}
or a shadow and a very thick eyebrow.
In this particular BBH,
the effect of the individual black hole spins on the image depends
strongly on the camera position.

\section{Conclusions}
\label{sec:conclusion}

In this paper, we present the first images of gravitational
lensing by astrophysically relevant binary black holes,
thereby providing a realistic representation of what an observer
near such a system would actually see.
To accomplish this, we have developed a new set of equations that evolve
photons efficiently near black hole horizons.
Our images show there is a primary shadow---a region where the black hole
prevents light from reaching the camera---for each black hole,
as well as multiple secondary shadows (or eyebrows).

We have found that,
early in the inspiral, images of a BBH look similar to
two separate Kerr black hole shadows, unless viewed when the holes are nearly
collinear with the camera.
Shortly before the merger, all camera angles yield interesting images of not
just one shadow for each black hole, but a handful of smaller visible shadows.
We showed for an equal-mass binary viewed edge-on that the lensing structure
exhibits self-similarity on smaller scales,
corresponding to photons taking
an increasing number of orbits through the system.
Lensing by a fully generic BBH illustrated that the spin
of black holes
in a binary can have a clear effect on the lensed shadows.

We chose not to classify eyebrows and shadows into a
hierarchy in this paper.
In the inset of figure~\ref{fig:EqualMassTop},
for instance, identifying
the largest eyebrow as the primary eyebrow and the next largest as the
secondary eyebrow feels very natural, but the exact
definition of such a
hierarchy is not immediately clear.
For example, simply specifying a geodesic winding number
around each black hole is likely not to be sufficient.
In addition to the trajectories not lying in a plane,
the order that a geodesic orbits the black holes
does not commute.
Furthermore, the black holes are moving at comparable speeds to
the geodesics.
For these reasons, we leave the task of classifying shadows as future work.

We have also shown in this paper that, away from the shadows, an image of a
binary black hole system looks like that of an isolated black hole. Thus it is
necessary to resolve individual shadows in order to discern the unique visual
characteristics present in such images, which places limits on our ability to
observe them.

For systems involving matter, however, the combination of the lensing effects
of strong gravity with the disruption and distortion of radiation-emitting
matter might yield a unique optical signature.
Generating lensed images of
black hole-neutron star and neutron star-neutron star mergers
is an avenue of future investigation.
The techniques presented here would allow us to produce
detailed visualizations of these mergers;
integrating over such images, we could predict the optical signature of
an unresolved system.

%\vspace*{2ex}
\begin{acknowledgments}
We would like to thank Curran Muhlberger for providing the temperature fits to
the 2MASS photometric data.
This publication makes use of data products from the Two Micron All Sky Survey,
which is a joint project of the University of Massachusetts and the Infrared
Processing and Analysis Center/California Institute of Technology, funded by
the National Aeronautics and Space Administration and the National Science
Foundation.
We would like to thank Daniel Hemberger and Saul Teukolsky for comments on
an earlier version of this paper.
The authors from Cornell would also like to thank Saul Teukolsky and
Lawrence Kidder for general advice while writing this paper.

This work was supported in part by NSF Grants PHY-1306125 and AST-1333129
at Cornell University, by NSF Grants PHY-1440083, AST-1333520,
PHY-1005655, and DMS-1065438 at the California Institute of Technology,
and by a grant from the Sherman Fairchild Foundation.
FH acknowledges support by the NSF Graduate Research Fellowship under
Grant No. DGE-1144153.
DB acknowledges support from the LIGO Laboratory, with funding
from the National Science Foundation under cooperative agreement PHY-0757058
and NSF REU award PHY-1062293.
The binary black hole simulations were performed using the Zwicky computer
system operated by the Caltech Center for Advanced Computing Research and
funded by NSF MRI No. PHY-0960291 and the Sherman Fairchild Foundation.
\end{acknowledgments}

%\section*{References}  %only for iopart
\pagebreak
\bibliographystyle{hunsrt}
\bibliography{References/References}

\end{document}